\documentstyle[11pt,newpasp,twoside]{article}
\def \stis{{\it STIS}}
\def \soho{{\it SOHO}}
\def \hst{{\it HST}}

\def \lya{Lyman~$\alpha$}
\def \hi{H$\,\sc i$}
\def \hd{HD$\,$209458}
\def \hdb{HD$\,$209458b}

\def\edcomment#1{\iffalse\marginpar{\raggedright\sl#1\/}\else\relax\fi}
\reversemarginpar

\begin{document}
\title{The~data~analysis~of~HD~209458\\ \lya~transit~occultation}
 \author{J.-M.~D\'esert, A.~Vidal-Madjar, A.~Lecavelier des \'Etangs, G.~H\'ebrard,
G.~Ballester, R.~Ferlet, M.~Mayor } \affil{Institut
d'Astrophysique de Paris, CNRS and UPMC, 98 bis Boulevard Arago,
 F-75014 Paris}

\begin{abstract}
Using \stis\ - \hst\ observations, we have detected an extended
upper atmosphere in the extrasolar planet \hdb . The spectra show
an absorption of 15$\pm$4\% of the stellar \lya\ emission line
during the planetary transit in front of the star. This is
interpreted as the signature of  \hi\ evaporation taking place
around the planet within an extended upper atmosphere
(Vidal-Madjar et al., 2003). We present the data analysis of these
observations. A particular attention is paid to the background
correction and the airglow subtraction.
\end{abstract}

The transit of a giant extrasolar planet offers a unique
opportunity to investigate the spectral features of its
atmosphere. Three transits of \hd\ were sampled in 2001 with the
STIS spectrograph onboard the Hubble Space Telescope (HST). We
used the G140M grating with the 52"x0.1" slit. For each transit,
three consecutive HST orbits were scheduled such that the first
orbit ended before the first contact to serve as a reference, and
the two following ones were partly or entirely within the transit.

\

The background level was systematically increasing in the
two-dimensional (2D) images from one exposure to the next within
each of the three visits but always remained below 2\% of the peak
intensity of the stellar signal. We therefore reprocessed the 2D
images by using two independent approaches.The resulting
difference between both approaches are found to be negligible,
showing that systematic errors generated through the background
corrections are small compared to the statistical errors.

The geocoronal emission filled the aperture of the spectrograph,
allowing us to remove it at the position of the target star. We
evaluated its variation both along the slit and from one exposure
to another. We concluded that the geocoronal contamination can be
removed with high enough accuracy outside the central region
(1215.5\AA $<$$\lambda$$<$1215.8\AA, named "Geo"). Because of
potential problems in  this domain, it is excluded from the
present analysis.

\

The three exposures outside the transits and the three entirely
within the transits were respectively co-added to improve the
signal to noise ratio. An obvious absorption in the hydrogen line
profile is detected during the transits, mainly over the blue side
of the line, and possibly at the top of the red peak (between -130
km s$^{-1}$ and +100 km s$^{-1}$). To characterize this signature
better, we have defined two spectral domains: "In" and "Out" of
the absorption. The "In" domain covers the interval
1215.15\AA-1216.1\AA\ excepted the "Geo" region. The "Out" domain
is the remaining wavelength coverage within the interval
1214.4\AA-1216.8\AA. During the transit, the corresponding "In" to
"Out" flux ratio appears significantly below 1. We derived that
the Lyman $\alpha$ line is reduced by 15$\pm$4\% (1$\sigma$).

\

We have investigated the possibility that a bad estimate of the
geocoronal correction might cause the signal we detected. the
change of the geocoronal correction by +30\% to -100\% (or even no
"Geo" correction) has a negligible impact on the evaluation of the
"In"/"Out" ratios; the absorption signature is still perfectly
visible in all cases and even when no correction at all is
applied. This indicates that the signal cannot be a consequence of
an improper geocoronal correction. To further show that the tail
of the geocorona does not carry the detected transit signal, we
looked at the "Geo"/"Out" signal for the uncorrected spectra. They
are random fluctuations around the average: there is no "transit"
signal in the geocoronal variations. We conclude that the
geocorona does not contaminate the transit signature seen in the
"In"/"Out" ratio (see Vidal-Madjar and Lecavelier, this book).

Since the HD209458 type is close to solar (G0V), we have evaluated
the "In" over "Out" ratio in the solar Lyman~$\alpha$ line profile
as measured by the SUMER instrument onboard \soho\ (Lemaire et al.
2002) during the last solar cycle from 1996 to 2001. During this
time, the total solar Lyman~$\alpha$ flux varies by about a factor
two, while its "In"/"Out" ratio varies by less than $\pm$6\%. This
suggests that the detected absorption is not of stellar origin but
rather due to a transient absorption occurring during the
planetary transits.

A bright hot spot on the stellar surface hidden during the
planetary transit is also excluded. Such a hot spot would have to
contribute about 15\% of the Lyman $\alpha$ flux over the 1.5\% of
the stellar surface occulted by the planet, in contradiction with
usual Lyman $\alpha$ inhomogeneities observed on the Sun (Prinz,
1974). Furthermore, this spot would have to be perfectly aligned
with the planet throughout the transit, at the same latitude as
the Earths direction, and with a peculiar narrow single-peaked
profile confined over the "In" spectral region.
 \

In summary, the observed 15\% Lyman $\alpha$ intensity drop is not
due to a geocoronal or stellar contamination but corresponds to an
occultation by an object of 4.3 Jupiter radii; this is clearly
beyond the Roche limit. Furthermore, the spectral absorption width
shows independently that the atoms have large velocities relative
to the planet. Thus hydrogen atoms must be escaping the planetary
atmosphere.

\vspace{0.4cm} \noindent {\bf References:}

\noindent Lemaire, P., Emerich, C. et al.
% , Vial, J.-C., Curdt, W., Schuhle, U., Wilhelm, K.
, 2002, ESA SP-508.

\noindent Prinz, D. K., 1974, ApJ. 187, 369-375.

\noindent
Vidal-Madjar, A., %Lecavelier des \'Etangs, A.,
et al.
%D{\' e}sert, J.-M., Ballester, G.~E., Ferlet,
%R., H{\' e}brard, G., \& Mayor, M.\
2003, Nature, 422, 143 (and these proceedings)

\vspace{0.3cm} \noindent {\small We thank M. Lemoine, L. Ben
Jaffel, C. Emerich, P. D. Feldman, M. K. Andr\'e and J. McConnell
for useful comments, J. Herbert and W. Landsman for useful
conversations on STIS data reduction, and J. Valenti for helping
in preparing the observations.}

\end{document}